 \documentclass[final,5p,times,twocolumn,authoryear]{elsarticle}

\usepackage{amssymb}
\usepackage{lipsum}
\usepackage{amsmath}
\usepackage{orcidlink}
\usepackage{booktabs}

\journal{Physics Letters B}
\begin{document}
\begin{frontmatter}

%
\title{Glitch noise classification in KAGRA O3GK observing data \\using unsupervised machine learning}
\author[first]{Shoichi Oshino\orcidlink{0000-0002-2794-6029}}
\affiliation[first]{organization={Institute for Cosmic Ray Research, KAGRA Observatory, The University of Tokyo},
            addressline={238 Higashi-Mozumi, Kamioka-cho}, 
            city={Hida City},
            postcode={506-1205}, 
            state={Gifu},
            country={Japan}}
\author[second]{Yusuke Sakai\orcidlink{0000-0001-8810-4813}}
\affiliation[second]{organization={Research Center for Space Science, Advanced Research Laboratories and Department of Design and Data Science, Tokyo City University},
            addressline={3-3-1 Ushikubo-Nishi, Tsuzuki-Ku}, 
            city={Yokohama},
            postcode={224-8551}, 
            state={Kanagawa },
            country={Japan}}
\author[second,second_second]{Marco Meyer-Conde\orcidlink{0000-0003-2230-6310}}
\affiliation[second_second]{University of Illinois at Urbana-Champaign, Department of Physics, Urbana, Illinois 61801-3080, USA}
\author[first]{Takashi Uchiyama\orcidlink{0000-0003-2148-1694}}
\author[third]{Yousuke Itoh\orcidlink{0000-0003-2694-8935}}
\affiliation[third]{organization={Graduate School of Science, Osaka Metropolitan University},
            addressline={ 3-3-138 Sugimoto-cho}, 
            city={Sumiyoshi-ku},
            postcode={558-8585}, 
            state={Osaka},
            country={Japan}}
\author[fourth,fourth_second,fourth_third]{Yutaka Shikano\orcidlink{0000-0003-2107-7536}}
\affiliation[fourth]{organization={Institute of Systems and Information Engineering, University of Tsukuba},
            addressline={1-1-1}, 
            city={Tennodai, Tsukuba},
            postcode={305-8573}, 
            state={Ibaraki},
            country={Japan}}
\affiliation[fourth_second]{organization={Center for Artificial Intelligence Research, University of Tsukuba},
            addressline={1-1-1}, 
            city={Tennodai, Tsukuba},
            postcode={305-8577}, 
            state={Ibaraki},
            country={Japan}}
\affiliation[fourth_third]{organization={Institute for Quantum Studies, Chapman University},
            addressline={1 University Dr.}, 
            city={Orange},
            postcode={92866}, 
            state={California},
            country={USA}}
\author[fifth]{Yoshikazu Terada\orcidlink{0000-0002-4509-1108}}
\affiliation[fifth]{organization={Graduate School of Engineering Science, The University of Osaka},
 addressline={1-3}, 
            city={Machikaneyama, Toyonaka},
            postcode={560-8531}, 
            state={Osaka},
            country={Japan}}
\author[second,sixth]{Hirotaka Takahashi\orcidlink{0000-0003-0596-4397}}
\affiliation[sixth]{organization={Earthquake Research Institute, The University of Tokyo},
 addressline={1-1-1}, 
            city={Yayoi, Bunkyo-ku},
            postcode={113-0032}, 
            state={Tokyo},
            country={Japan}}
\begin{abstract}
Gravitational wave interferometers are disrupted by various types of nonstationary noise, referred to as glitch noise, that affect data analysis and interferometer sensitivity. The accurate identification and classification of glitch noise are essential for improving the reliability of gravitational wave observations. In this study, we demonstrated the effectiveness of unsupervised machine learning for classifying images with nonstationary noise in the KAGRA O3GK data. Using a variational autoencoder (VAE) combined with spectral clustering, we identified eight distinct glitch noise categories. 
The latent variables obtained from VAE were dimensionally compressed, visualized in three-dimensional space, and classified using spectral clustering to better understand the glitch noise characteristics of KAGRA during the O3GK period. 
Our results highlight the potential of unsupervised learning for efficient glitch noise classification, which may in turn potentially facilitate interferometer upgrades and the development of future third-generation gravitational wave observatories.
\end{abstract}
%
%
%
\end{frontmatter}
%
\section{Introduction}\label{introduction}
KAGRA~\citep{2019NatAs...3...35K}, the next-generation gravitational wave detector, conducted its first observation run in collaboration with GEO600~\citep{Willke_2004} from 8:00 UTC on April 7, 2020 to 0:00 UTC on April 21, 2020 (O3GK)~\citep{03GK_paper}. 
The collected data are publicly available as open data~\citep{O3GK_open_data_1,O3GK_open_data_2}~\footnote{https://gwosc.org/O3/O3GK/ (Accessed August 2025)}.

Since then, LIGO-Virgo-KAGRA has established an international joint observation network for conducting gravitational wave observations~\citep{GWTC-1,GWTC-2,GWTC-2.1,GWTC-3}. 
During these collaborative observing runs, various environmental and instrumental transients, such as ground vibration, lightning, pendulum control signals, and laser fluctuations, affect the interferometer and become mixed with the gravitational wave data. 
These nonstationary and non-Gaussian noises are termed as `glitch' noises. LIGO and Virgo collaborations reported that glitch noises with a signal-to-noise ratio $> 6.5$ occurred at a rate of $1.10$ events per minute at LIGO Livingston (LLO) in the first half of the third observation run (O3a) between April 1, 2019, 15:00 UTC and October 1, 2019, 15:00 UTC~\cite{GWTC-2}, and at a rate of $1.17$ events per minute at LLO in the second half of O3 (O3b) between November 1, 2019, 15:00 UTC and  March 27, 2020, 17:00 UTC~\cite{GWTC-3}.

Detecting and classifying glitch noise is an important step from the following perspectives: 
\begin{enumerate}
\item[1)]Glitch detection techniques enable the separation of glitch noise from gravitational waves originating from astronomical phenomena,
\item[2)]Glitch classification techniques aid in identifying the sources of glitch noise,
\item[3)]Identifying the sources of glitch noise facilitates their removal and increases the amount of data available for analysis and improvement of interferometer sensitivity.
\end{enumerate}
Hence, the development of a project, termed Gravity Spy~\citep{2017CQGra..34f4003Z, BAHAADINI2018172, 2024EPJP..139..100Z}, was initiated.
Gravity Spy project involves citizen scientists to assign labels to training data, enabling the classification of glitch noises detected by LIGO into 22 distinct types. This supervised learning model achieves high accuracy in automatically classifying glitch noise.

Efforts have been made to understand the glitch noise in KAGRA by analyzing data from O3GK \citep{kihyun}.
However, the interferometer configurations of KAGRA and LIGO are different, and even in the case of LIGO, new glitch noise can appear as the interferometer is upgraded. 
Additionally, several challenges exist in applying the same approach as the Gravity Spy project to KAGRA: unlike the Gravity Spy project, KAGRA does not have citizen scientists to assist in manual classification and labeling.
Furthermore, the sensitivities of KAGRA and LIGO interferometers are different; therefore, the appearance of identical glitch noises is not guaranteed. 
To resolve these difficulties, in this study, we first focus on the classification of glitch noise in KAGRA's O3GK data using unsupervised learning methods.

\cite{2022NatSR..12.9935S, 2024AnP...53600140S} explored the effectiveness of unsupervised learning algorithms using data from the Gravity Spy project. 
Their study used a variational autoencoder (VAE) \citep{2013arXiv1312.6114K, MAL-056} for feature learning to extract latent variables from the time-frequency spectrogram image of glitch noise.
The extracted latent variables were visualized in a three-dimensional (3D) space using dimensionality compression via Uniform Manifold Approximation and Projection (UMAP)~\citep{2018arXiv180203426M}.
In the study, potential improvements over Gravity Spy were observed, and new glitch noise shapes that were not manually classified were proposed~\citep{2022NatSR..12.9935S, 2024AnP...53600140S}.  

In our study, we used VAE architecture to extract the latent variables from the time--frequency spectrogram image of the glitch noise in O3GK.
We applied UMAP to the extracted latent variables to visualize the clustering of glitch noises.
Finally, we classified the glitch noise using spectral clustering~\citep{von2007tutorial} in the visualized space and determined the multiple classes in O3GK data.

The remainder of this paper is organized as follows. 
Section~\ref{sec:framework} provides a brief overview of the dataset, preprocessing steps, and analysis methods.
The results and discussions are presented in Section~\ref{sec:Result}.
Finally, Section~\ref{sec:summary} summarizes our findings.

\section{Proposed Framework}\label{sec:framework}
\begin{figure*}
	\centering 
	\includegraphics[width=0.7\textwidth]{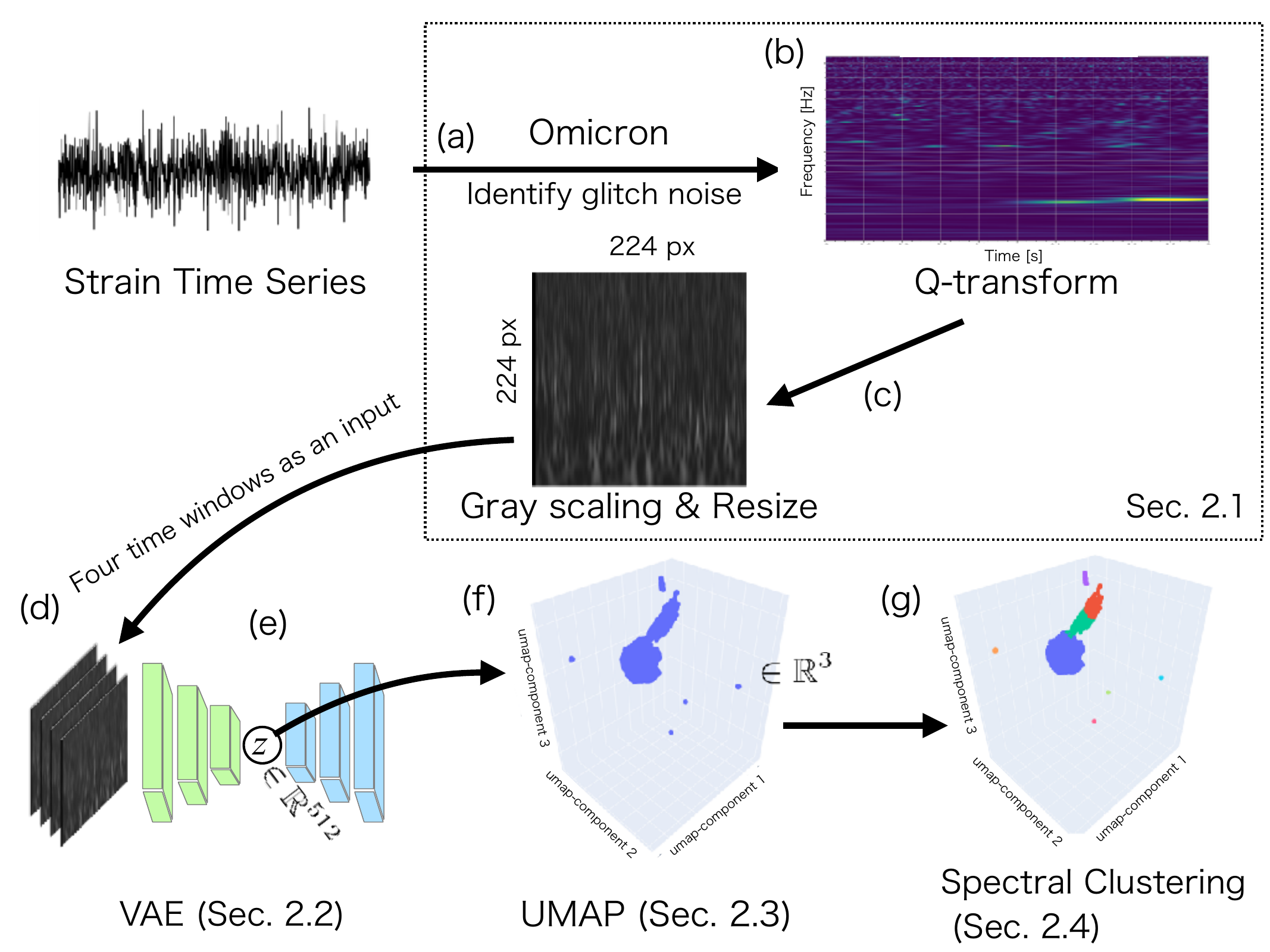}	
	\caption{Data analysis flow in this study. (a) Datastream acquired by the KAGRA interferometer were processed using the Omicron pipeline to identify glitching times. (b) Creation of a time-frequency spectrogram for each trigger. (c) Conversion of the spectrograms to grayscale. (d) Integration of four time ranges into a composite image. (e) Training the created dataset VAE to extract latent variables. (f) Dimensional compression into 3D space using UMAP. (g) Glitch noise shape classification using spectral clustering.} 
	\label{fig:procedure-overview}
\end{figure*}

In this study, we classified the glitch noise in the O3GK data using the following procedure:
\begin{enumerate}[a)]
    \item Datastream acquired by the KAGRA interferometer were processed using the Omicron pipeline to identify glitching times.
    \item A time--frequency spectrogram was created for each trigger.
    \item The spectrograms were converted to grayscale as a preliminary step in preparing data for machine learning.
    \item Four time ranges were integrated into a composite image.
    \item A VAE was trained on the created dataset to extract latent variables.
    \item Subsequently, the distribution of glitch noises was ascertained by dimensional compression into 3D space using UMAP.
    \item Finally, glitch noise shapes were classified using spectral clustering.
\end{enumerate}
The aforementioned procedure is illustrated in Fig.~\ref{fig:procedure-overview}. The details are as follows.

\subsection{Datasets}
We first generated glitch noise images using the O3GK KAGRA dataset, which is approximately 178 h of data. For this purpose, we used the Omicron software~\citep{2020SoftX..1200620R} to identify transient noise events and compile a database of their GPS timestamps from the strain data.
Omicron is an event trigger generator, which can be used to produce spectrograms from whitened data streams and search for transient detector noise and gravitational wave events. In this study, we used the same configuration of O3 ~\footnote{https://github.com/gw-detchar/tools/blob/master/Omicron/\\Parameter/O3rerun\_C20.txt (Accessed August 2025)}
We selected data with peak frequencies ranging from 10 \rm{Hz} to 2048 \rm{Hz} and a signal-to-noise ratio exceeding $7.5$ based on the Gravity Spy Project.
Under these conditions, the detected glitch noise was $4.63$ events per minute.
The GPS times were then applied to the Omega Scan pipeline \citep{2004CQGra..21S1809C} to create Q-transformed images, which created time--frequency spectrograms from time-series data by setting a window function for each time--frequency period.
As with the Gravity Spy dataset, we created a set of images for four time windows (0.5, 1.0, 2.0, and 4.0 s) from the center time of the glitch noise event.
A single glitch noise image of size 224 \rm{px} $\times$ 224 \rm{px} was rescaled from the originally generated image of size 800 \rm{px} $\times$ 600 \rm{px}. 

Four such time-window images were stacked to obtain glitch noise data with a shape of 4 $\times$ 224 \rm{px} $\times$ 224 \rm{px} (bottom left, Fig.~\ref{fig:procedure-overview}). The image data were finally converted from color to grayscale for training datasets.
In this study, Omicron detected 45,345 glitch noises.

\subsection{Variational Autoencoder}
Generative models capture useful features from the input data and facilitate effective understanding of the underlying structure.
A VAE is a generative model consisting of an encoder, which transforms an input into a latent variable (denoted by $z$), and a decoder that reconstructs the input from 
this latent representation.
The model is trained by optimizing the Evidence Lower BOund (ELBO), which comprises a reconstruction term and regularization term that enforce the latent variable distribution to approximate a prior distribution,  a standard Gaussian. Various extensions have been proposed to expand the choice of prior distributions, such as VampPrior~\citep{tomczak2018vae} and Normalizing Flows~\citep{rezende2015variational}, which enhance expressiveness and flexibility in modeling complex data distributions.

In this study, we employed a VAE architecture~\citep{2022NatSR..12.9935S, 2024AnP...53600140S}, which comprises convolutional neural network (CNN) layers, as listed in Table~\ref{tab:vae_architecture}.
This architecture has previously demonstrated the capability of unsupervised clustering of glitch noises in the Gravity Spy project. To explore the generalization ability of this architecture across different datasets, we applied it to O3GK glitch noise and investigated its effectiveness in clustering.
While clustering is performed using features compressed into a three-dimensional latent space, higher-dimensional representations, typically $\leq$10 dimensions, are often necessary to produce sufficient expressiveness~\citep{ref:saha2025ard}.
Therefore, we also adopted and investigated a higher-dimensional latent representation in our model.

\begin{table}[t]
  \centering
  \caption{Architecture of VAE comprises sequential layers.
  $d_z$ denotes the dimension of the latent variable, $C_{\text{out}}$ denotes the number of output channels of the feature, and $k_{\text{s}}$, $s$, and $p$ represent the kernel size, stride, and padding, respectively.
  For further details, see~\citep{2022NatSR..12.9935S, 2024AnP...53600140S}.
  }
  \label{tab:vae_architecture}
  \begin{tabular}{@{}ll@{}}
    \toprule
    Module & Description \\
    \toprule
    Encoder & 
    \begin{tabular}[t]{@{}l@{}}
    \\
    \midrule
    Definition of EncoderBlock $(C_{\text{out}}$, $k_{\text{s}}$, $s$, $p$):\\
    \quad - 2D convolutional layer $(C_{\text{out}}$, $k_{\text{s}}$, $s$, $p$)\\
    \quad - Batch normalization\\
    \quad - ReLU activation map\\
    \midrule
    Input: 4-channel image of shape $(4, 224, 224)$ \\
    EncoderBlock($64$, $k_{\text{s}}=7$, $s=2$, $p=3$) \\
    Max-pooling layer with $k_{\text{s}}=3$, $s=2$, $p=1$. \\
    EncoderBlock($128$, $k_{\text{s}}=3$, $s=2$, $p=1$) \\
    EncoderBlock($256$, $k_{\text{s}}=3$, $s=2$, $p=1$) \\
    EncoderBlock($512$, $k_{\text{s}}=3$, $s=2$, $p=1$) \\
    \end{tabular} \\
    
    & 
    \begin{tabular}[t]{@{}l@{}}
    Adaptive average pooling layer \\
    Linear layer: outputs $z \in \mathbb{R}^{d_z}$ \\
    \end{tabular} \\
    \midrule
    
    Decoder & 
    \begin{tabular}[t]{@{}l@{}}
    \\
    \midrule
    Definition of DecoderBlock $(C_{\text{out}}, k_{\text{s}}, s, p)$: \\
    \quad - Upsampling by nearest-neighbor interpolation \\
    \quad - 2D convolutional layer $(C_{\text{out}}, k_{\text{s}}, s, p)$ \\
    \quad - Batch normalization \\
    \quad - ReLU activation map\\
    \midrule
    Input: latent variable $z$\\
    Linear layer: $\mathbb{R}^{d_z} \to \mathbb{R}^{d_z \times 7 \times 7}$ \\
    Batch-normalization\\
    ReLU activation \\
    Upsampling by nearest-neighbor interpolation \\
    DecoderBlock($256$, $k_{\text{s}}=3$, $s=2$, $p=0$) \\
    DecoderBlock($128$, $k_{\text{s}}=3$, $s=2$, $p=0$) \\
    DecoderBlock($64$, $k_{\text{s}}=3$, $s=2$, $p=0$) \\
    DecoderBlock($4$, $k_{\text{s}}=3$, $s=2$, $p=0$)\\
    \end{tabular} \\
    \bottomrule
  \end{tabular}
\end{table}

\subsection{Uniform Manifold Approximation and Projection}
UMAP~\citep{2018arXiv180203426M} is a nonlinear dimensionality reduction method that provides low-dimensional embedding while preserving the distance relationships in high-dimensional data. 
Compared with t-SNE (\cite{roweis2000nonlinear}), UMAP produces visualizations more rapidly and captures the underlying cluster structure in the data more clearly.
Specifically, UMAP enables the embedding of newly observed data points into existing low-dimensional representation without the need to recompute the entire embedding.
This functionality renders UMAP well-suited for real-time analysis, with potential applications such as glitch noise clustering in live data streams.
It is widely used in various fields as a state-of-the-art method for dimensionality reduction and visualization.

UMAP algorithm \citep{2018arXiv180203426M} uses a specific distance metric to compute an undirected weighted $k$-nearest neighbor graph based on the input data.
Subsequently, it embeds this neighbor graph into a low-dimensional space while minimizing a specific loss function called the fuzzy loss entropy, which is related to the embedded and input data.
UMAP applications~\footnote{https://github.com/lmcinnes/umap (Accessed August 2025)} require a tuning parameter $\delta$, which affects the compactness of the clusters in the embedded space. A larger value of $\delta$ generates a more spread-out cluster structure in the resulting embedding.
In accordance with our previous investigation, we set the Euclidean distance as the distance metric, $k=10$ as the number of neighbors, and $\delta = 0.05$ for UMAP.

\subsection{Spectral Clustering}
Clustering is the process of grouping data points such that points within the same group (\textit{i.e., } cluster) are more similar to each other than to those in different groups.
It can be categorized into supervised clustering with labels and unsupervised clustering without labels. The well-established algorithms $k$-means++~\citep{arthur2006k} and spectral clustering~\citep{von2007tutorial} are categorized as unsupervised clustering. Among these, spectral clustering is better for handling complex data such as non-convex data distributions.

In the spectral clustering algorithm~\footnote{https://github.com/scikit-learn/scikit-learn/blob/main/sklearn/cluster (Accessed August 2025)}, the adjacency matrix of a graph within the input data is computed. The adjacency matrix represents the similarity between data points and determines the normalized graph cuts~\citep{shi2000normalized}.
In general, the following Gaussian kernel is often used to compute the adjacency matrix $A = (a_{ij})_{n\times n}$:
\begin{equation}
    a_{ij} = \exp\left(
    -\frac{||x_i - x_j||^2}{2\sigma^2}
    \right),
    \label{eq:gaussian-kernel}
\end{equation}
where $n$, $x_i$, and  $||\cdot||^2$ denote the number of data samples, $i$th data vector in the input, and  Euclidean distance, respectively, while 
$\sigma^2$ represents a tuning parameter of Gaussian kernel.
The median heuristic~\citep{garreau2017large} is often used to select $\sigma^2$ because of its effective performance.
The formula for the median heuristic $\sigma^2_{\mathrm{MH}}$ is expressed as
\begin{equation}
\sigma^2_{\mathrm{MH}} \stackrel{\mathrm{def}}{=} \mathrm{Median}
        \{||x_i = x_j||^2 \mid 1 \leq i < j \leq n\}.
        \label{eq:median-heuristic}
\end{equation}

We classify the latent variable of the glitch noise using spectral clustering with the median heuristic such that $x_i$ is replaced by the latent variable $z_i$ in Eq.~\eqref{eq:gaussian-kernel} and Eq.~\eqref{eq:median-heuristic}: 
Given that the number of classes in O3GK glitch noise is unknown, we used the Davies--Bouldin index (DBI) \citep{4766909} to determine the optimal number of classes in addition to a careful visual inspection of the spectral shapes of the glitches. Further details are presented in the following section. 

\section{Results and Discussions}\label{sec:Result}
The hyperparameters of VAE, which were investigated based on a previous study~\citep{2022NatSR..12.9935S, 2024AnP...53600140S}, are as follows:
The dimensions of the latent variable $z$ are $\{32, 64, 128, 256, 512\}$; the minibatch size is $\{32, 64, 96, 128\}$; the number of epochs is 100; the learning rate is $5\times10^{-4}$, using the Adam optimizer~\citep{kingma2014adam}; and the ratio of training data and test data is $80\%\colon 20\%$.
The training curves showed no significant differences compared to those in previous studies, suggesting that the VAE architecture remained stable across a range of parameter settings, even when applied to different glitch noise datasets.
In this study, the dimensions of latent variable $z$ and minibatch size were $512$ and 96, respectively.

\begin{figure*}
	\centering 
	\includegraphics[width=0.7\textwidth]{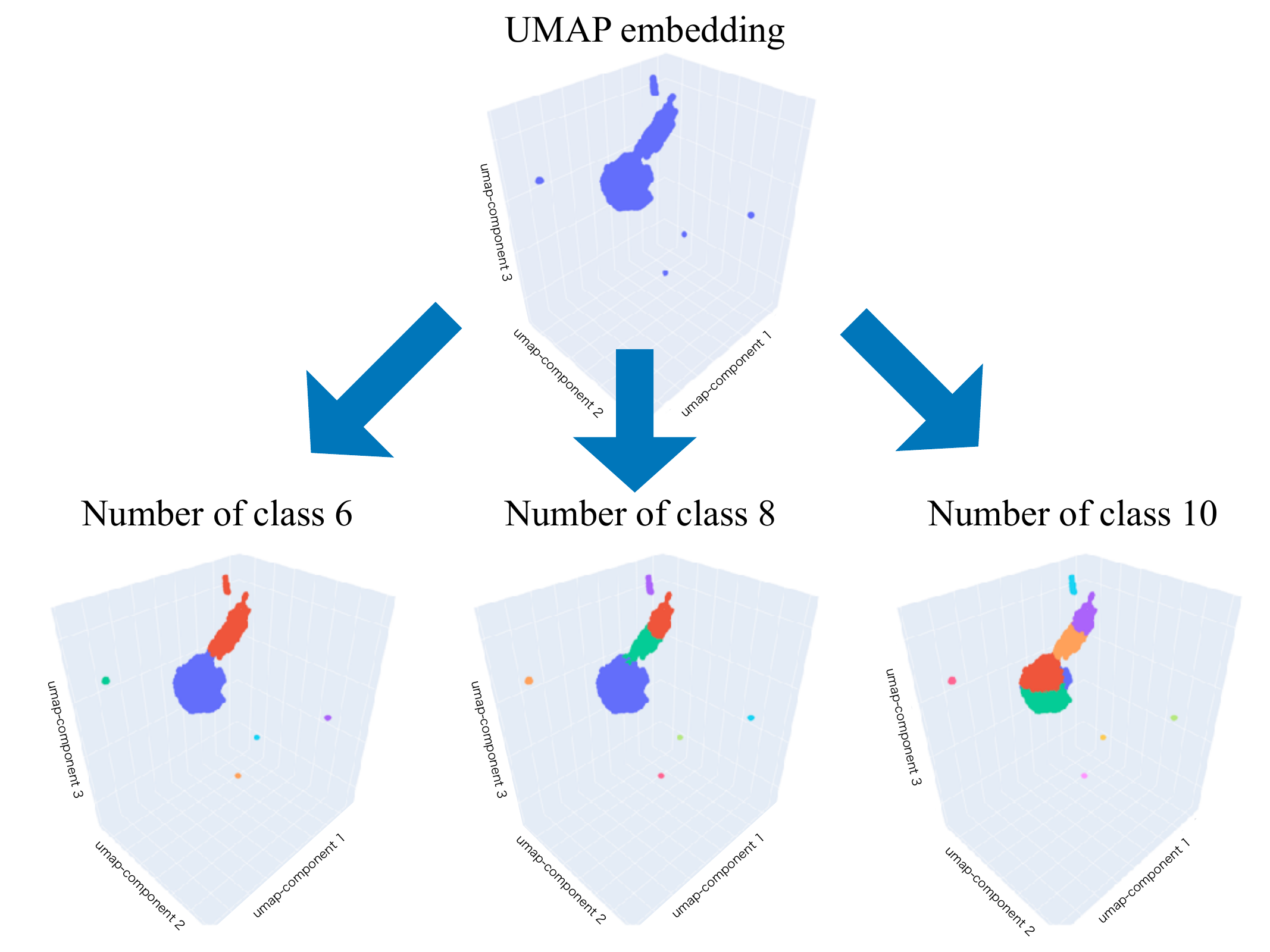}	
	\caption{Glitch noise visualization in 3D space using UMAP with full O3GK data. Additionally, the figures show glitch noise classified using spectral clustering and color-coded by class. The difference between the color-coded figures is the difference in the number of classes divided by spectral clustering. From left to right in the bottom row, the figures are classified into 6, 8, and 10 classes.} 
	\label{fig:UMAP}
\end{figure*}

To verify the distribution of features in the latent variables that were trained using the VAE, dimensional reduction was performed in 3D space using UMAP. 
The upper part of Fig.~\ref{fig:UMAP} shows the embedded latent variables.
The visualization of the latent variables shows that O3GK glitch noise has a cluster structure, which comprises several small clusters and one or two large clusters.

To classify glitch noise, we performed spectral clustering on the embedded latent variables.
The number of glitch noise shapes in the KAGRA O3GK dataset was classified as 6~\citep{kihyun}, which is fewer than the 22 labeled in the Gravity Spy dataset. Spectral clustering was performed with the number of classes ranging from four to 12.
The lower part of Fig.~\ref{fig:UMAP} presents the results of the classification of the embedded latent variables.
The numbers of class divisions (6, 8, and 10) specified by spectral clustering are color-coded from left to right in the lower row.
With a smaller division number, such as 6, large and small clusters are grouped together, whereas a larger division number, such as 10, results in further subdivision of large clusters.

\begin{figure}
	\centering 
	\includegraphics[width=0.4\textwidth]{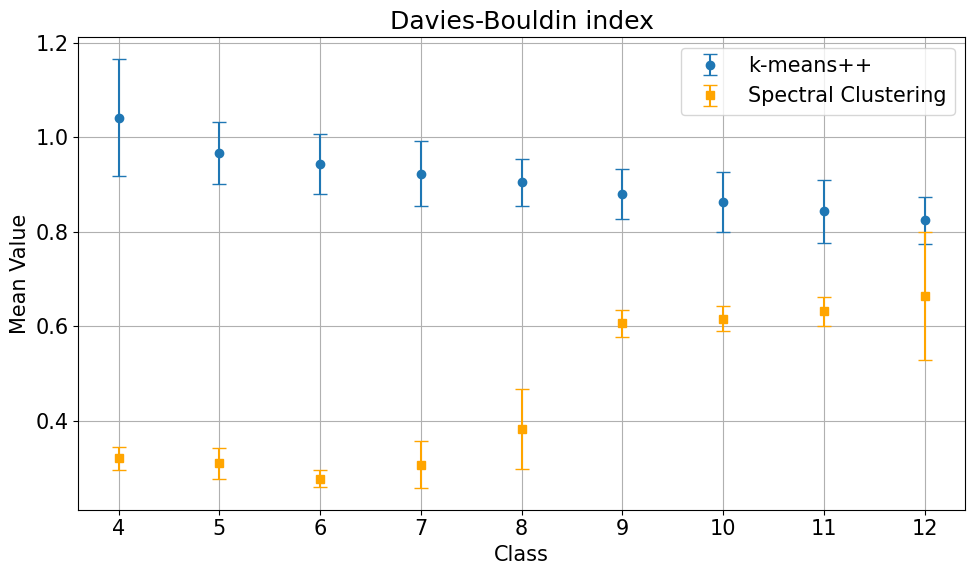}	
	\caption{ Average Davies--Bouldin index values for $k$-means++ and spectral clustering as a function of the number of clusters. Each clustering experiment was repeated 20 times with different random seeds to account for stochastic variability, and the mean index was computed. Error bars represent the standard error of the mean. Blue and orange are the results of $k$-means++ and spectral clustering, respectively.} 
	\label{fig:dbi}
\end{figure}

In~\ref{sec:appendix}, we investigated classification using $k$-means++ and determined that it was less suitable for O3GK clustering than for spectral clustering.
The DBI~\citep{4766909}, which compares the distance between clusters with the size of the clusters themselves, was used as an index of the number of clusters. 
A value close to zero indicates better partitioning. 
The results of the DBI computed using spectral clustering and $k$-means++ for each division are presented in Fig~\ref{fig:dbi}, 
which shows the mean and standard deviation of 20 calculations. For spectral clustering, the score increases after eight classes, whereas for $k$-means++, the DBI decreases as the number of classes increases.

As a complement to the DBI evaluation, the results of spectral clustering were also evaluated using the silhouette coefficient~\citep{rousseeuw1987silhouettes}, as discussed in~\ref{sec:appendix_silhouette}.
These results are consistent with the results shown in Fig~\ref{fig:dbi}, which were
evaluated using the DBI.

\begin{figure*}
	\centering 
	\includegraphics[width=0.9\textwidth]{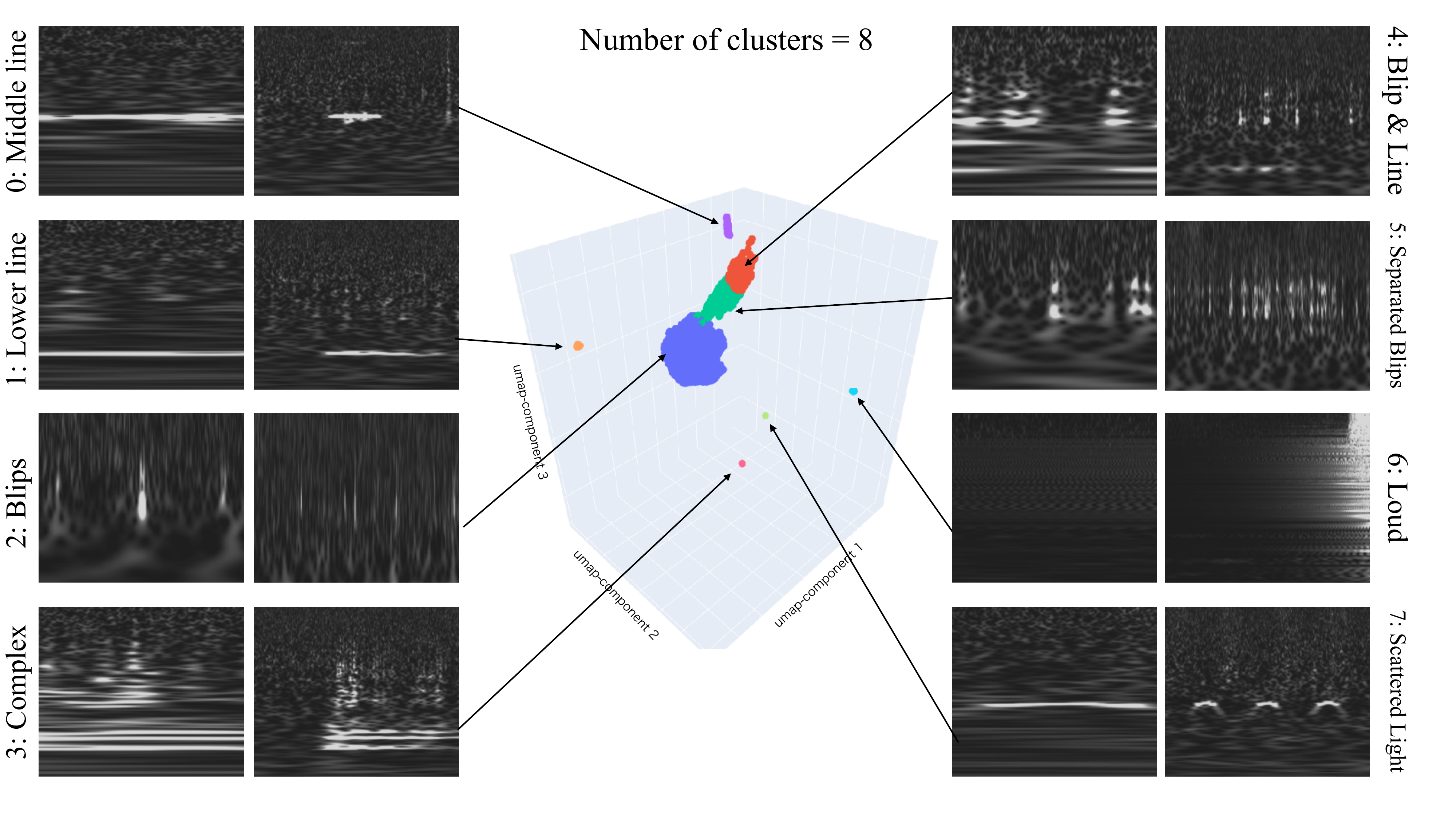}	
	\caption{Latent variables of full O3GK data were dimensionally compressed into a 3D space using UMAP and then classified. Each glitch noise image illustrates the characteristic glitch noise shape for each cluster. The glitch noise image has time windows of 0.5 s and 4 s on the left and right sides, respectively, for each class.} 
	\label{fig:SC}
\end{figure*}

In Fig.~\ref{fig:SC}, the color-coded latent variables on UMAP are displayed from the results of spectral clustering when the number of clusters is eight.
Each cluster is accompanied by a representative noise image, which spans time windows of 0.5 s (left) and 4 s (right).
Based on the DBI, the optimal number of clusters for spectral clustering was between four and eight. However, inspection of the glitch noise images in Fig.~\ref{fig:SC} shows that images \#2, \#4, and \#5, which were all assigned to the largest cluster, exhibited distinctly different shapes. Therefore, considering both the DBI results and visual observations, we conclude that eight is the most appropriate number of clusters in the spectral clustering.

Table~\ref{tab:sc_class} shows that the number of glitch noises in each cluster for the number of class divisions is eight.
The most common glitch noise is class \#2, which accounts for approximately $80 \%$ of total noise.
The noise shape of this class is teardrop-shaped and resembles the shapes termed as ``Blips'' in the Gravity Spy dataset.
The next largest clusters are classes \#4 and \#5, which have vertical noise shapes that are wider than that of class \#2.
Additionally,  a lateral linear structure appears while transitioning from class \#5 to class \#4.
Therefore, class \#5 is termed ``Separated Blips'' because it resembles a ``Blip'' that has been split.
Class \#4 is denoted as ``Blip $\&$ Line'' because horizontal lines appear in addition to vertical lines.
Classes \#0 and \#1 exhibit strong lateral linear structures.
Class \#0, termed as the ``Middle line'', has a line positioned centrally along the vertical direction, while class \#1, referred to as the ``Lower line'', has a line at the bottom.
Class \#3 is denoted ``Complex'' because it includes various shapes.
Class \#7 is ``Scattered Light'', which is also used in the Gravity Spy dataset.
Therefore, we confirmed that glitch noises of this shape were present in the photodetector data    
 when ``Scattered Light'' glitch noises were observed in the strain data during the KAGRA O3GK period.
Finally, the last class, class \#6, is ``Loud'' because it is a loud noise.

The types of glitch noises detected in the KAGRA O3GK dataset are fewer than those labeled in LIGO by the Gravity Spy project. 
This discrepancy may be attributed to the sensitivity of the KAGRA interferometer during the O3GK period, which was significantly lower than that of LIGO.
Thus, the noise floor was high for the transient glitch noise; these may have been screened by the stationary noise.
Therefore, higher sensitivity of KAGRA is expected during O4, which may generate new glitch noise shapes. Even in this case, new glitch noise shapes can be easily identified by applying the proposed method to O4 data.


\begin{table}[tb]
    \centering
\caption{Results of glitch noise classification when spectral clustering is executed with 8 classes. The typical glitch noise images are described in Fig~\ref{fig:SC}. The name of the glitch noise is with reference to the glitch noise image of Gravity Spy.} \label{tab:sc_class}
    \begin{tabular}{c r l}
    \hline
    Class & Number of glitch noise & Shape of glitch noise\\
    \hline
    0 & 621 (1.4\%) & Middle line \\
    1 & 294 (0.6\%) & Lower line \\
    2 & 35925 (79.2\%) & Blips \\
    3 & 44 (0.1\%) & Complex \\
    4 & 4016 (8.9\%) & Blip \& Line \\
    5 & 4358 (9.6\%) & Separated Blips \\
    6 & 60 (1.3\%) & Loud \\
    7 & 27 (0.6\%) & Scattered Light \\
    \hline
    \end{tabular}
\end{table}

\section{Summary}\label{sec:summary}
Gravitational wave observation data streams contain numerous transient noises that hinder data analysis and the improvement of interferometer performance. Classifying these transient noises is essential for understanding their sources and in advancing gravitational wave science.

We confirmed that unsupervised learning (VAE) can be effectively applied to KAGRA data. 
The latent variables obtained from VAE were dimensionally compressed using UMAP, visualized in 3D space, and classified using spectral clustering to effectively understand the glitch noise characteristics of KAGRA during the O3GK period. 

In the future, the same method will be applied to data from the current O4 observations to investigate the evolution of the topology of glitch noises in response to changes in the interferometer configuration of KAGRA.

\section*{Acknowledgments}
This research was partly supported by Japan Society for the Promotion of Science (JSPS) Grant-in-Aid for JSPS Fellows [No. 22KF0329 (M.~Meyer-Conde)] and Grants-in-Aid for Scientific Research [Nos. 23H01176, 23K25872 ,and 23H04520 (H.~Takahashi)].
This research was supported by the Joint Research Program of the Institute for Cosmic Ray Research, University of Tokyo and Tokyo City University Prioritized Studies.

\appendix
\section{Classification using $k$-means++}\label{sec:appendix}
\begin{figure}
	\centering 
	\includegraphics[width=0.4\textwidth]{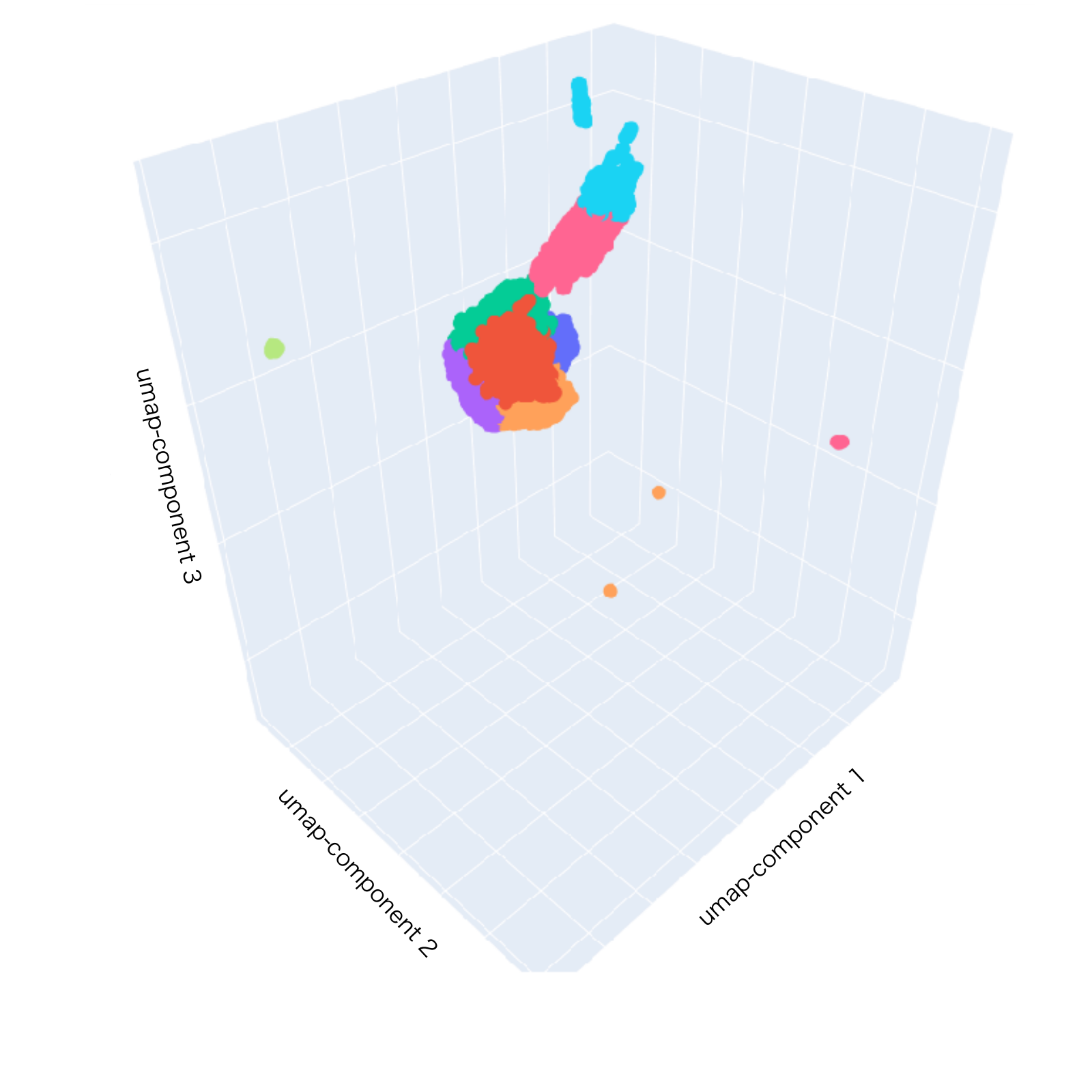}	
	\caption{Latent variables of full O3GK data plotted in a 3D space using UMAP and divided and color-coded using $k$-means++. The number of divisions is eight.}
	\label{fig:kmeans_8}
\end{figure}

The $k$-means algorithm is a standard clustering method that assigns each data point based on its distance from the center of mass of a cluster.
$k$-means++~\citep{arthur2006k} is a variant of the $k$-means algorithm that achieves better clustering by strategically initializing the cluster centers.
We used $k$-means++ to classify O3GK glitch noise on the embedded latent variable (Fig~\ref{fig:kmeans_8}).
The $k$-means++ method prioritizes the dense clumps in the latent variable space for partitioning. In contrast, scattered latent variables are grouped into the same class.
Owing to these characteristics, $k$-means++ prioritizes the division of the cluster that has the largest number.

Examining and comparing clustering methods is essential for determining the most suitable approach for detecting glitch noise in data obtained from laser interferometric gravitational wave detectors.
As discussed in Section~\ref{sec:Result}, spectral clustering performed well on the O3GK, highlighting the importance of evaluating multiple methods for identifying the optimal solution.

\section{Silhouette Coefficient}\label{sec:appendix_silhouette}
As a complement to the DBI evaluation, Fig.~\ref{fig:dbi}, discussed in the main text, the results of spectral clustering were evaluated using the silhouette coefficient~\citep{rousseeuw1987silhouettes}.
Although the DBI assesses the separation and compactness of clusters, the silhouette coefficient quantifies the degree to which each sample conforms to its assigned cluster.
This coefficient approaches 1 when the clustering is compact and well-separated, and it approaches 0 when the number of clusters is not appropriate. A negative value indicates that the data points may overlap with other clusters.

\begin{figure}
	\centering 
	\includegraphics[width=0.4\textwidth]{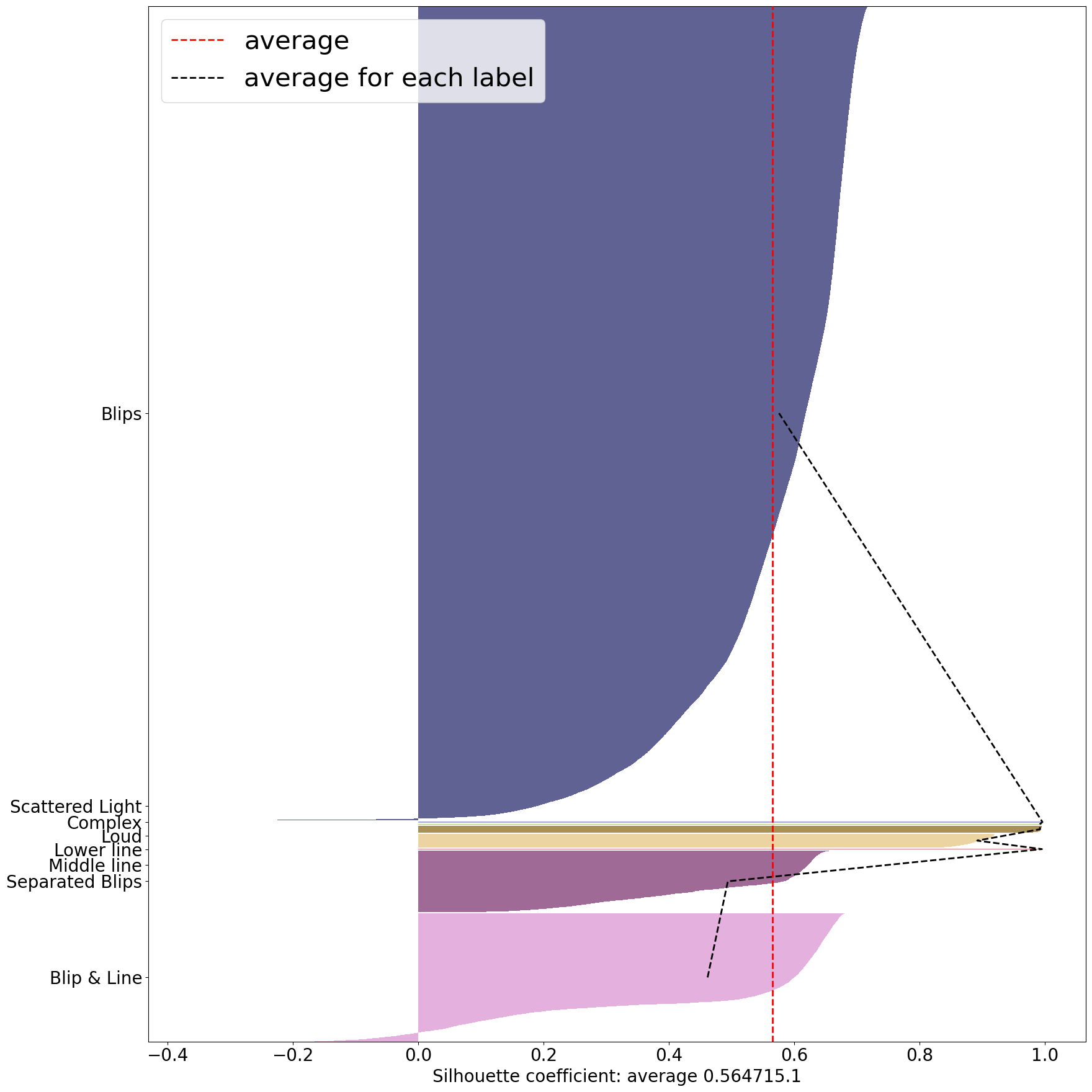}	
	\caption{Distribution of silhouette coefficient for all O3GK data when the number of classes is 8. The black dashed line indicates the average silhouette coefficient for each label, while the red dashed line shows the overall average across all labels.}
	\label{fig:silhouette}
\end{figure}

Fig.~\ref{fig:silhouette} shows the silhouette coefficients of each glitch noise when divided into eight classes using spectral clustering. Overall, the silhouette coefficients are positive, indicating that the clustering was condensed.

\begin{figure}
	\centering 
	\includegraphics[width=0.4\textwidth]{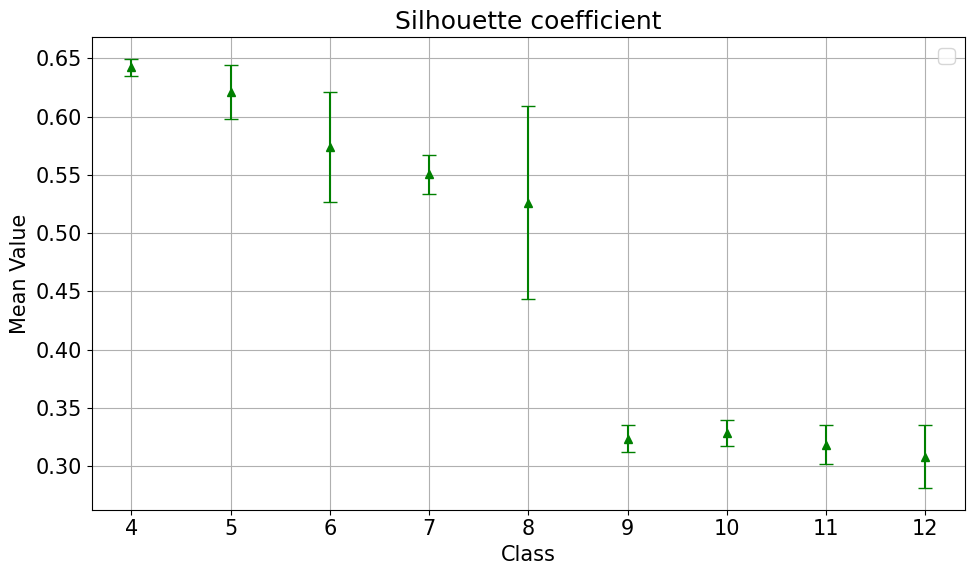}	
	\caption{Results for calculating the silhouette coefficients for spectral clustering when the number of classes is changed. The silhouette coefficient, originally computed as a vector for each sample, was averaged to obtain a scalar value per trial.}
	\label{fig:silhouette_class}
\end{figure}

Fig.~\ref{fig:silhouette_class} shows the results of calculating the silhouette coefficients for spectral clustering when the number of classes is changed. These individual silhouette coefficients have been averaged across all samples to obtain a scalar value. To assess the stability and reproducibility of the clustering method, the clustering process was repeated 20 times using different random seeds. For each iteration, the average silhouette coefficient was computed; these values were then averaged again. Finally, the mean and variance for each number of classes were plotted. This figure demonstrates that increasing the number of clusters beyond eight leads to a decline in clustering compactness. This result is consistent with the results shown in Fig.~\ref{fig:dbi}, which were evaluated using the DBI.

%
\bibliographystyle{elsarticle-harv} 
\bibliography{references}

\end{document}